\documentclass{article}

\usepackage{mathptmx}
\usepackage{arxiv}
\usepackage{graphicx}
\usepackage{hyperref}      
\usepackage{amsmath,amsfonts,amssymb}
\usepackage{color}
\usepackage{braket}
\usepackage{float}
\usepackage{subfig}
\usepackage{cite}

\hypersetup{
    colorlinks=true,
    linkcolor=red,
    filecolor=cyan,      
    urlcolor=magenta,
    citecolor=blue}

\title{Statistical properties of London modified coherent states}
\author{H\'ector M. Moya-Cessa
\\ Instituto Nacional de Astrofísica Óptica y Electrónica\\Calle Luis Enrique Erro No. 1, Santa María Tonantzintla, Pue., 72840, Mexico.
  \\ \And Julio Guerrero\\
 Departamento de Matem\'aticas, Universidad de Ja\'en, Campus las Lagunillas, 23071 Ja\'en, Spain.\\
}

\begin{document}
\maketitle

\begin{abstract}
In this paper we discuss statistical properties of modified London coherent states, that were
introduced in order build up a resolution of the identity for London coherent states. In particular, we show that there exist sub-Poissonian behaviour for a large interval of amplitudes and, because their oscillating photon distribution, they show ringing revivals of the atomic inversion.

\end{abstract}


\section{Introduction}
\label{intro}
There are various definitions of generalized or nonlinear  coherent states \cite{Manko,RasettiI,RasettiII,RasettiIII,Recamier,Ancheyta,Dodonov} for a given system: Perelomov states \cite{PerelomovI,PerelomovII}, that are the application of a displacement operator to the vacuum; Barut-Girardello states \cite{Barut}, that are eigenstates of the (nonlinear) annihilation operator;  coherent states that have a resolution to the identity \cite{KlauderI,KlauderII}; and minimum uncertainty states \cite{Manko,Vaccaro,Trifonov}. In particular, Glauber coherent states \cite{Glauber,Sudarshan} obey these four properties, however, to the best of our knowledge, there are no other states that comply with those four definitions simultaneously.

Recently, Le\'on-Montiel {\it et al.} \cite{LeonI,LeonII} have introduced what they wrongly named Susskind-Glogower coherent states (as the name London is more adequate).  Those states are Perelomov states since they are obtained by the application of the exponential of the sum of the  operators
\begin{equation}
    V=\sum_{n=0}^{\infty} |n\rangle\langle n+1|, \qquad  V^{\dagger}=\sum_{n=0}^{\infty} |n+1\rangle\langle n|, \label{PhaseOp}
\end{equation}
to the vacuum. States $|n\rangle$ are number states, i.e., eigenstates of the harmonic oscillator.

These operators are usually known as Susskind-Glogower \cite{Susskind} operators, but they were introduced firstly by London \cite{London} as  candidates to proper phase operators of the form $V=exp(i \Phi)$, since the angle operator $\Phi$ was shown to be non-hermitean (the same year Dirac \cite{Dirac} also constructs  these operators, although quoting London's paper). Even Louisell \cite{Louisell} introduced the same operators (in the version of sine and cosine of the angle operator) a year before Reference \cite{Susskind} was published.

London coherent states may be written as  \cite{LeonI,LeonII}
\begin{equation}
   |x\rangle_L =e^{x(V-  V^{\dagger})}=\frac{1}{x}\sum_{n=0}^{\infty} (n+1)J_{n+1}(2x)|n\rangle,
   \label{LondonCs}
\end{equation}
where $x$ is the amplitude of the nonlinear coherent states\footnote{We call them nonlinear coherent states as they are related not to the creation and annihilation operators of the harmonic oscillator, but to generalizations of them containing functions of the number operator.}, which, for simplicity, we take as real (see Appendix for the extension to the complex plane of these states).

Note that Sudarshan \cite{Sudarshanhar} introduced Barut-Girardello-type coherent states for the
phase operators (\ref{PhaseOp}), denoted as Harmonius states, parametrized by complex numbers in the unit disk. These states are not related to the ones discussed in this paper and will not be further discussed.

London operators (\ref{PhaseOp}) have the problem that they are neither hermitean nor unitary. For these reasons there has been an active search for a hermitean phase operator and nowadays the more accepted proposal is that of
Pegg and Barnett \cite{PeggBarnett}, although other problems appear due to the regularization
 required for a proper definition of these operators and the associated phase states.
 Coherent states for  Pegg and Barnett have also been constructed \cite{PeggBarnettCS}.

Distributions of classical light that model London coherent states may be generated by propagating an electromagnetic field in inhomogeneous media, particularly, waveguide arrays \cite{Leija1,Observation}.  Those states are an infinite superposition of number states with coefficients that are proportional to Bessel functions.

One problem with these states is the difficulty in building a resolution of the identity (which is related to the fact that phase operators (\ref{PhaseOp}) do not close under commutation into a finite dimensional Lie algebra), see \cite{preparation} where the authors introduce and discuss modified London states.

Gazeau  \cite{Gazeau} has modified those states to comply with the resolution to the identity. The modified states read as
\begin{equation}
   |x\rangle_{Lm} =\frac{1}{N(x)}\sum_{n=0}^{\infty} \sqrt{n+1}J_{n+1}(2x)|n\rangle,\label{modified}
\end{equation}
with $N(x)$ a normalization constant given by \cite{Gazeau}
\begin{equation}
    N(x)=\frac{1}{x^2}\sum_{n=1}^{\infty}nJ_n^2(2x).\label{normal}
\end{equation}

In this contribution we will study the statistical properties of the states defined in (\ref{modified}), their interaction with a two-level atom, that presents ringing revivals and will show how they may be obtained from the application of an exponential operator to the vacuum.
\section{Application of a generalized displacement operator to the vacuum}
First we note that the normalization (\ref{normal}) implies the interesting identity
\begin{equation}
   \sum_{n=1}^{\infty}nJ_n^2(x)=\frac{x^2}{2}[J_0^2(x)+J_1^2(x)]-\frac{x}{2}J_0(x)J_1(x),
\end{equation}
where we have used some techniques introduced by Dattoli \cite{DattoliI,DattoliII}, so we may write it as
\begin{equation}
   N(x)=2[J_0^2(2x)+J_1^2(2x)]-\frac{1}{x}J_0(2x)J_1(2x).
\end{equation}
Next we rewrite (\ref{modified}) in the form
\begin{equation}
   |x\rangle_{Lm} =\frac{x}{N(x)\sqrt{\hat{n}+1}}\sum_{n=0}^{\infty} \frac{n+1}{x}J_{n+1}(2x)|n\rangle,
\end{equation}
so that we may relate the London coherent states (from Perelomov definition) to the modified London coherent states (from the resolution of the identity) as
\begin{equation}
   |x\rangle_{Lm} =\frac{x}{N(x)\sqrt{\hat{n}+1}}|x\rangle_L, \label{relation}
\end{equation}
with $\hat{n}=a^{\dagger}a$, the number operator and $a$ and $a^{\dagger}$ the annihilation and creation operators of the harmonic oscillator.

Le\'on-Montiel {\it et al.} showed that $|x\rangle_L$ may be obtained from the application of a generalized displacement operator to the vacuum, therefore we may write
\begin{equation}
   |x\rangle_{Lm} =\frac{x}{N(x)\sqrt{\hat{n}+1}}e^{x(V-V^{\dagger})}|0\rangle. \label{displacement}
\end{equation}
We can further write (\ref{displacement}) by means of a nonunitary transformation
\begin{equation}
   |x\rangle_{Lm} =\frac{x}{N(x)}e^{-\ln \sqrt{\hat{n}+1}}e^{x(V-V^{\dagger})}e^{\ln \sqrt{\hat{n}+1}}|0\rangle,
\end{equation}
that, after some algebra has the final form
\begin{equation}
   |x\rangle_{Lm} =\frac{x}{N(x)}e^{x\left(\sqrt{\frac{{\hat{n}+2}}{\hat{n}+1}}V-V^{\dagger}\sqrt{\frac{{\hat{n}+1}}{\hat{n}+2}}\right)}|0\rangle.
\end{equation}
The above equation is related to non-Hermitian quantum mechanics  \cite{Bender} as, taking $x\rightarrow t$, it represents the (normalized) evolution operator of the non-Hermitian Hamiltonian
\begin{equation}
   H_{n-H}=i{\left(\sqrt{\frac{{\hat{n}+2}}{\hat{n}+1}}V-V^{\dagger}\sqrt{\frac{{\hat{n}+1}}{\hat{n}+2}}\right)}.
\end{equation}
\begin{figure}[htbp]
    \includegraphics{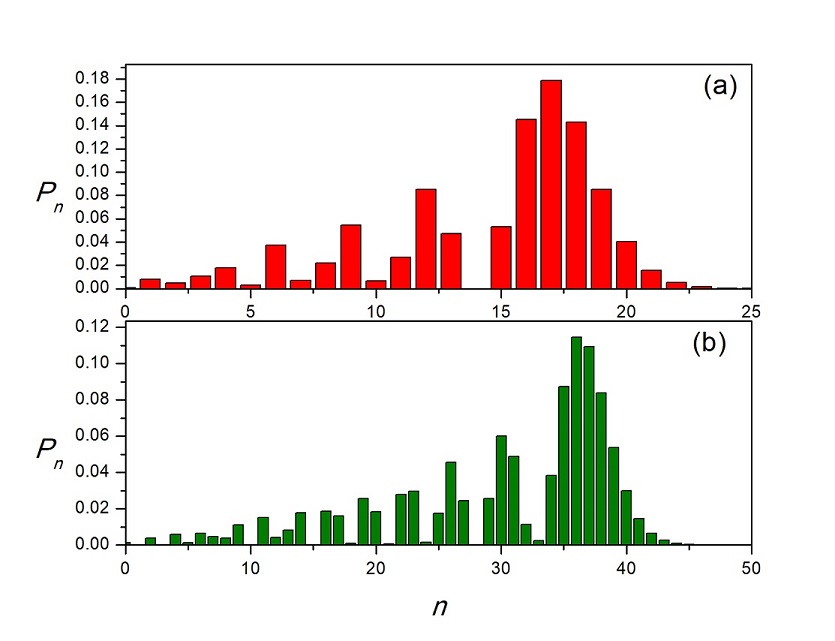}
    \caption{We plot the photon distribution of the modified London state for the amplitudes (a) $x=10$ and (b) $x=20$. } \label{Fig_2}
\end{figure}

It should be stressed, however, that modified London states (\ref{modified}) are not equivalent to London states (\ref{LondonCs}), due to the presence of the operator $\frac{1}{\sqrt{\hat{n}+1}}$ in their relation (\ref{relation}). In fact, the operator  $\frac{1}{\sqrt{\hat{n}+1}}$ is invertible but its inverse is not bounded. This implies that states $\sum a_n|n\rangle$ that are normalizable might be no longer normalizable
when transformed by the inverse of $\frac{1}{\sqrt{\hat{n}+1}}$. Also, Statistical properties of both family of states would be slightly different. In the next section we discuss the statistical properties of modified London coherent states (see \cite{LeonII} for the statistical properties of London coherent states).
\begin{figure}[htbp]
    \includegraphics{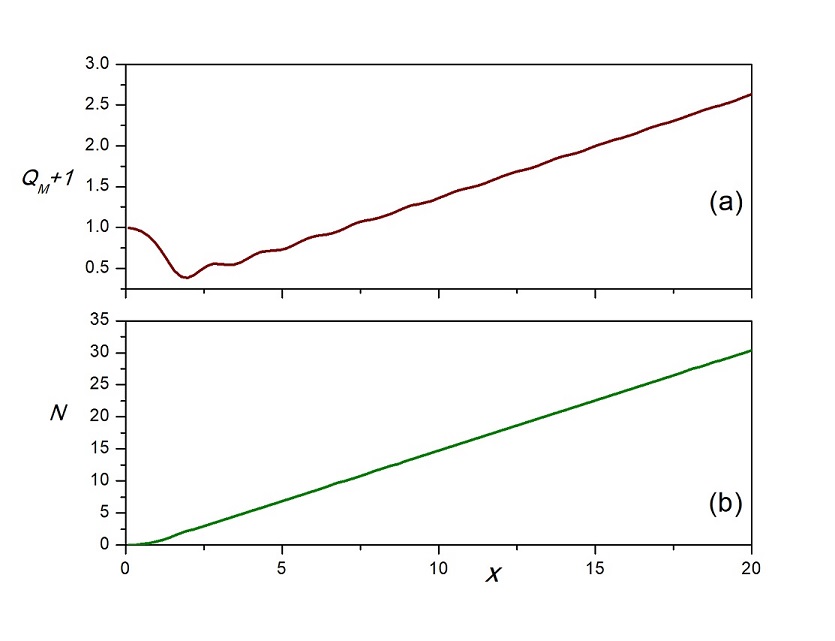}
    \caption{We plot (a) the Mandel $Q$-parameter (plus one) and (b) the average number of photons as a function of the amplitude $x$.  } \label{Fig_1}
\end{figure}
\section{Statistical properties of the modified London states}
\label{sec:1}
In this Section we show some statistical properties of the modified London states. From equation (\ref{modified}) we may obtain the photon distribution
\begin{equation}
   P_n=|\langle n|x\rangle_{Lm}|^2 =\frac{1}{N^2(x)} (n+1)J^2_{n+1}(2x).\label{distribution}
\end{equation}
We plot this photon distribution as  a function of the number of photons in Fig. 1.  It clearly shows oscillations in the photon distribution, that, as we will show in the next Section, lead to so-called ringing revivals of the atomic inversion.

In order to calculate the Mandel $Q$-parameter
\begin{equation}
  Q_M=\frac{_{Lm}\langle x|\hat{n}^2|x\rangle_{Lm}-N^2}{N}-1
\end{equation}
with  $N=_{Lm}\langle x|\hat{n}|x\rangle_{Lm}$ the average number of photons, that is  obtained from (\ref{distribution}) as
\begin{equation}
 N=\frac{1}{N^2(x)}\sum_{n=0}^{\infty} (n^2+n)J^2_{n+1}(2x),\label{average}
\end{equation}
and the second ingredient needed is the second moment, {\it i.e.},
\begin{equation}
 _{Lm}\langle x|\hat{n}^2|x\rangle_{Lm}=\frac{1}{N^2(x)}\sum_{n=0}^{\infty} (n^3+n^2)J^2_{n+1}(2x).\label{average}
\end{equation}

\begin{figure}[htbp]
    \includegraphics{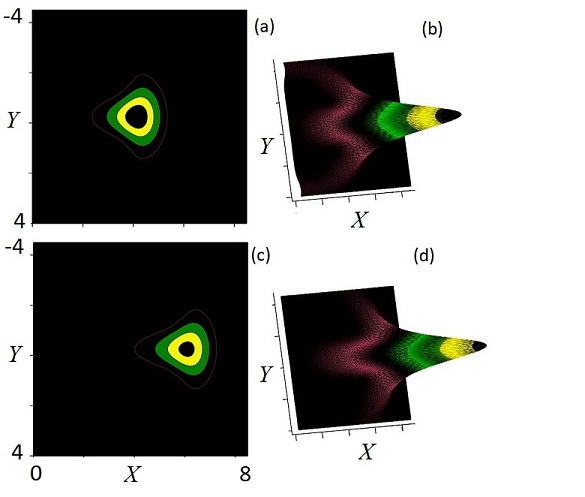}
    \caption{$Q$-function for the modified London state for $x=10$ [(a) and (b)] with a maximum value $Q_{max}\approx 0.24$, and for $x=20$ [(c) and (d)] with a maximum  value at $Q_{max}\approx 0.21$. } \label{Fig_3}
\end{figure}
We plot in Fig. 2 these two quantities, the Mandel $Q$-parameter showing a sub-Poissonian behaviour for amplitudes up to $x\approx 6$ while the average number of photons increases almost linearly with the amplitude.

Finally we show the Husimi $Q$-function
\begin{equation}
    Q(\alpha)=\frac{|\langle\alpha|x\rangle_{Lm}|^2}{\pi}
\end{equation}
a quasiprobability often used to visualize the system in phase space. Here $|\alpha\rangle$ is a coherent state. In Fig. 3 we plot it for [(a) and (b)] $x=10$ and for [(c) and (d)] $x=20$, with $\alpha=X+iY$.

\section{Interaction with a two-level atom}
\begin{figure}[htbp]
    \includegraphics{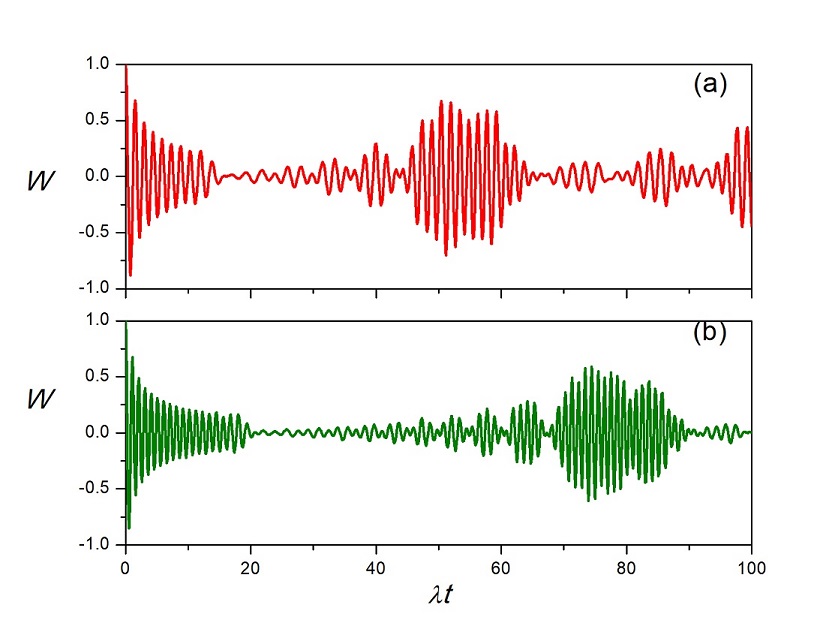}
    \caption{ We plot the atomic inversion as a funtion of time for an initial  modified London state with amplitudes (a) $x=10$ and (b) $x=20$. } \label{Fig_4}
\end{figure}
It is well-known that the atomic inversion (probability to find the atom in the excited state minus the probability to find it in the ground state) in the atom-field interaction (when the field is considered quantized and the atom with two levels) \cite{Jaynes} depends strongly on the behaviour of the photon distribution of the field \cite{Shore,Nonlinear,Casanova2018}. The atomic inversion for the resonant interaction between the quantized field and the two-level-atom is given by the expression
\begin{equation}
    W(t) =\sum_{m=0}^{\infty} P_m cos(\lambda t\sqrt{m+1}),
\end{equation}
where $\lambda$ is the atom-field interaction constant. In Fig. (\ref{Fig_4}) we show that the atomic inversion exhibits ringing revival behaviour \cite{Satyanarayana,Optik} generated by the photon distribution  plotted in Fig.  (\ref{Fig_1}).

\section{Conclusions}
We have studied the statistical properties of the London-modified states introduced by Gazeau as coherent states that obey the resolution to the identity. In particular we have shown that they may be obtained as the application of an displacement-like  operator to the vacuum. We have also studied their statistical properties by calculating the Mandel-$Q$ parameter which, for a large range of amplitudes, shows sub-Poissonian nature. We have also shown the appearance of ringing revivals of the atomic inversion, product of the oscillations of the photon distribution of the state.

\appendix
\section{}
In this appendix we generalize to complex values the London state
\begin{equation}
   |x\rangle_L =e^{x(V-  V^{\dagger})}|0\rangle
\end{equation}
This is done by a simple unitary transformation $|z\rangle_L=e^{-i\theta \hat{n}}|x\rangle_L$, which gives
\begin{equation}
   |z\rangle_L =e^{-i\theta \hat{n}}e^{x(V-  V^{\dagger})}|0\rangle=e^{(zV-  z^*V^{\dagger})}|0\rangle,
\end{equation}
with $z=xe^{i\theta}$ and where we have used the commutator $[\hat{n},V]=-V$.
%
% For  figures use
%\begin{figure*}
%\vspace*{5cm}       % Give the correct figure height in cm
%\includegraphics{leer.eps}
%\end{figure*}
% or  this
%\begin{figure}
%\centering
%\resizebox{0.75\textwidth}{!}{%
%  \includegraphics{leer.eps}
%\vspace{5cm}       % Give the correct figure height in cm
%\end{figure}
%\begin{table}
%\centering
%\begin{tabular}{lll}
%\hline\noalign{\smallskip}
%\end{tabular}
% Or use
%\vspace*{5cm}  % with the correct table height
%\end{table}
\bigskip

%
% BibTeX users please use
% \bibliographystyle{}
% \bibliography{}
{\bf Author Contributions:} Both authors conceived the idea and developed it. The manuscript was written by both
authors, who have read and approved the final manuscript.

\bigskip
{\bf Funding:} J.G. acknowledges financial support from the Spanish Ministerio de Ciencia e InnovaciÃ³n (PGC2018-097831-B-I00 and FIS2017-84440-C2-2-P). J.G. also thanks INAOE for financial support and hospitality during two short visits in 2015.
\bigskip

{\bf Conflicts of Interest:} The authors declare no conflict of interest.

\end{document}